\documentclass[preprint,5p,twocolumn,twoside,sort&compress]{elsarticle}

\usepackage{graphicx}
\usepackage[figuresright]{rotating}

\journal{Physica D}

\begin{document}

\begin{frontmatter}
\title{Simple model of cell crawling}

\author[toyota]{T. Ohta\corref{col1}}
\author[kyoto]{M. Tarama}
\author[tokyo]{M. Sano}

\cortext[col1]{Corresponding author. email:ohta@daisy.phys.s.u-tokyo.ac.jp}

\address[toyota]{Department of Physics,The University of Tokyo, Tokyo, 606-8502 Japan, and \\
Toyota Physical and Chemical Research Institute, Nagakute, Aichi 480-1192, Japan}
\address[kyoto]{Fukui Institute for Fundamental Chemistry, Kyoto University, Kyoto, 606-8103 Japan}
\address[tokyo]{Department of Physics,The University of Tokyo, Tokyo, 606-8502 Japan}

\begin{abstract}
Based on symmetry consideration of migration and shape deformations, we formulate phenomenologically  the dynamics of cell crawling in two dimensions. Forces are introduced 
to change the cell shape. The shape deformations induce migration of the cell on a substrate. For time-independent forces 
we show that not only a stationary motion but also a limit cycle oscillation of the migration velocity and the shape occurs as a result of nonlinear coupling between different deformation modes. 
Time-dependent forces are
generated in a stochastic manner  by utilizing the so-called coherence resonance of an excitable system. 
The present coarse-grained model has a flexibility that it can be applied, e.g.,  both  to keratocyte cells and to ${\it Dictyostelium}$ cells, which exhibit quite different dynamics from each other. The key factors for the motile behavior inherent in each  cell type are identified in our model. 
\end{abstract}


\end{frontmatter}

%%%%%%%%%%%%%%%%%%%%%%
\section{Introduction}

Eukaryotic cell crawling has attracted much attention recently from the view point of nonlinear science and non-equilibrium statistical physics. One of the characteristic features  is that the symmetry is spontaneously broken to cause the  front-rear asymmetry when the cell migrates in contrast to bacterias which swim by rotary motion of flagella and hence are inherently asymmetric. The dynamics of eukaryotic cells involves the mechanical forces  between  cell membrane and substrate, and biochemical reaction of active molecules inside a cell. 

Study of cell crawling on substrates has began rather recently compared to that of swimming bacterias. The latter has a long history of hydrodynamical approach in the limit of low Reynolds number \cite{Blake, Purcell, Shapere, Farutin}.  Shape deformation of crawling keratocyte cells has been analyzed experimentally \cite{Keren}. Classification of morphology of motile cells, correlations between shape deformation and migration of ${\it Dictyostelium}$ cells and other living cells have also been investigated \cite{Li, Bosgraaf, Maeda, Tanaka}. Recent advanced experimental techniques have enabled us to measure the spatial distribution of traction forces exerted by a migrating cell on substrates \cite{Wang, Fournier, Style, Tanimoto} and the concentration  distribution of active molecules which involve  cell motility \cite{Taniguchi}. 

Plasma membrane protrusion caused by actin polymerization in the cell interior is the essential mechanism of cell crawling. In the early 1990's, DiMilla et al investigated  persistent migration of tissue cells such as fibroblasts by a mathematical model which is essentially one-dimensional and incorporates cytoskeletal force generation, cell polarization, and dynamic adhesion \cite{DiMIla}. Theoretical studies of cell crawling taking into account shape deformations have started recently. 
Modeling of cell crawling employing reaction-diffusion mechanism inside a cell or on a cell boundary and the interaction between the chemical components and the cell membrane has been proposed in two dimensions \cite{Keshet, Levine, Satulovsky}. A phase field model for cell shape  coupled with the polarization field of actin filaments has also been proposed for  the cell motility.
The crawling dynamics of keratocyte cells including oscillatory straight motion and bipedal motion \cite{Barnhart} has been investigated  by taking account of the density of adhesion bonds and traction forces \cite{Ziebert1, Ziebert2}. A similar model in terms of the phase field has been studied in which the reaction-diffusion dynamics  inside a ${\it Dictyostelium}$ cell is assumed to be excitable and therefore this model is capable of investigating non-stationary motion  \cite{Taniguchi}. Modeling of amoeboidal cell crawling has also been formulated by an oscillatory dynamics  \cite{Carlsson2} where
irregularity appears as spatio-temporal chaos  \cite{Dreher}. These are models in two dimensions. 
In a slightly different approach, theory of active gels has been applied to stationary amoeba motion in one dimension to make a connection between the migration velocity and the distribution of active stress or myosin molecules \cite{Carlsson, Recho}.   
Motility of active droplets in which active stress is generated has been studied numerically both in two and three dimensions to show that a bifurcation  from a motionless state to a migrating state with shape deformations occurs due to spontaneous symmetry breaking of the polarity inversion in the absence of treadmilling \cite{Tihung}.

It is mentioned briefly that  there is another mechanism of cell motility due to plasma membrane blebbing \cite{Fackler}. This is not restricted to migration on substrate. Blebbing is initiated by local disruption of membrane-action cortex and internal hydrostatic pressure. It is of importance to note that actin polymerization is not involved in the initial bleb expansion.  Cell motility by blebbing in three-dimensional environments has  been investigated theoretically \cite{Hawkins}.

In the present paper, we study cell crawling under a homogeneous environment based on a phenomenological model in terms of migration velocity and shape deformations. A set of time-evolution equations is derived based on symmetry consideration in the same spirit as the derivation of equations for deformable self-propelled particles \cite{Ohta, Hiraiwa}.   
To make shape deformations, we introduce forces which act on the cell perimeter.
We consider time-independent and time-dependent forces separately. 
The case of constant forces is regarded as a model of coherent motions, e.g., of keratocyte cells whereas the time-dependent forces are applied to motility of ${\it Dictyostelium}$ cells. 
 In the experiments of  ${\it Dictyostelium}$ cells, morphological change occurs repeatedly but it is not precisely periodic. To realize this behavior, we utilize the so-called coherence resonance which generates spike excitation of chemical components repeatedly in an excitable system when noise is added appropriately \cite{Coherence}.

All the  previous models mentioned above are constructed to apply to a specific system such as fish keratocyte cells and ${\it Dictyostelium}$ cells. A steadily migrating keratocyte cell is elongated perpendicularly to the velocity direction \cite{Keren} whereas a ${\it Dictyostelium}$ cell in a starved condition has a tendency to elongate parallel to the migration direction \cite{Maeda}. Our model, though simple,  has an advantageous feature that it is applicable to  non-stationary motion of a crawling cell with general shape deformations by choosing appropriately the parameters. 

One of the basic assumptions  of our model in terms of the center of mass and the cell boundary is that all other degrees freedom involving migration relax rapidly. If this is not the case, we need to add other relevant degrees of freedom as dynamical variables. 

In the next section (section 2) we start with description of our model system. The case of constant forces is analyzed in section 3 where we obtain limit cycle oscillations of migration velocity and shape deformations. This is compared with the phase field model of keratocyte cells. Numerical results for the time-dependent forces are shown in section 4 and are compared qualitatively with the motions of ${\it Dictyostelium}$ cells. Discussion is given in section 5.

%%%%%%%%%%%%%%%%%%%%%%
\section{Model for cell crawling} \label{section:model}

We introduce the model of cell crawling in two dimensions in terms of the velocity of the center of mass $v_i$, and the deformation tensors
\begin{eqnarray}
 v_{k} &=&  \gamma S_{ij}U_{ijk}  ,
 \label{eq1} \\
\frac{d S_{ij}}{dt}&=&-\kappa_2 S_{ij} +b_0(v_iv_j-\frac{\delta_{ij}}{2}v_kv_k)+F_{ij}^{(2)}(t)   ,
 \label{eq2}  \\
  \frac{d U_{ijk}}{dt}&=& -\kappa_3 U_{ijk}+d_0[v_iv_jv_k  \nonumber \\
  &-&\frac{v_nv_n}{4}( \delta_{ij}v_k+\delta_{jk}v_i+\delta_{ki}v_j)] + F_{ijk}^{(3)}(t)  ,
\label{eq3}
\end{eqnarray}
where the repeated indices imply summation.  

The tensors $S_{ij}$ and $U_{ijk}$ are defined as follows. 
Deformations of a cell around a circular shape with radius $R_0$ are written  as 
\begin{equation}
R(\phi, t) = R_0 (1+ \delta R(\phi, t) )
 \;,
\label{eq:cn}
\end{equation}
where
\begin{eqnarray}
 \delta R(\phi, t)  = \sum_{n=-\infty}^{\infty} c_n(t) e^{i n \phi}
  \;.
\label{eq:deltaR}
\end{eqnarray}
Since uniform expansion and contraction of a circular cell are prohibited and the translational motion of the cell has been incorporated in the variable $v_k$, the modes $c_{0}$ and $c_{\pm 1}$ should be removed from the Fourier series (\ref{eq:deltaR}).
The deformation tensors are given in terms of the Fourier coefficients  by \cite{OOS}
\begin{eqnarray}
 S_{11} &=& c_{2}+ c_{-2}=2a_2\cos2\theta_2  \label{S11}  , \\
 S_{12} &=&S_{21} = i(c_{2}- c_{-2})=2a_2\sin2\theta_2  \label{S12}  ,  \\
  S_{22} &=&-S_{11}  ,
 \label{S22} 
\end{eqnarray}
\begin{eqnarray}
U_{111}=-U_{122}=-U_{212}=-U_{221}\equiv W_+  \label{U111}   , \\
U_{222}=-U_{112}=-U_{121}=-U_{211}\equiv -W_-  ,
 \label{U222} 
\end{eqnarray}
where 
\begin{eqnarray}
 W_{+} = c_{3}+ c_{-3}=2a_3\cos3\theta_3  \label{W+}  , \\
 W_{-} = i(c_{3}- c_{-3})=2a_3\sin3\theta_3   ,
 \label{W-} 
\end{eqnarray}
with positive $a_2$ and $a_3$.

The coefficients $\kappa_2$ and $\kappa_3$ are positive while the sign of $\gamma$ in eq. (\ref{eq1}) will be fixed later. 
Here, for simplicity,  we ignore other nonlinear couplings  such as $U_{ijk}v_k$ and $S_{ij}v_k+S_{jk}v_i+S_{ki}v_j-(v_n/2)(\delta_{ij}S_{kn}+ \delta_{jk}S_{in}+ \delta_{ki}S_{jn})$ \cite{Ohta} but consider the coupling only with the velocity as eqs. (\ref{eq2}) and (\ref{eq3}) since those terms are expected to be mostly relevant to the correlation between the elongation direction and the migration direction as shown below. 

Equation (\ref{eq1}) implies that there is no inertia term and that the cell does  not  migrate if it is circular since we consider a deformation-induced migration.
In our previous studies of  migration-induced deformations \cite{OOS,Yabunaka}, equation of motion of the center of mass was derived, which takes the following form
\begin{eqnarray}
\frac{d v_k}{dt}=\kappa_1 v_{k}-g (v_i)^2v_k+ a S_{kj}v_j+\gamma' S_{ij}U_{ijk}  ,
 \label{eq4}
\end{eqnarray}
 where $\kappa_1$, $g(>0)$, $a$ and $\gamma'$ are constants. Equation (\ref{eq1}) is a special case of eq. (\ref{eq4}). In fact, when $\kappa_1$ is negative, that is, migration is passive, one may ignore the $g$-term and in the limit of $|\kappa_1| \gg 1$, the solution of eq. (\ref{eq4}) is given by
 \begin{eqnarray}
 v_k=-\gamma'[\kappa_1+ a S]^{-1}_{k\ell} S_{ij}U_{ij\ell} \approx -\frac{\gamma'}{\kappa_1}S_{ij}U_{ijk}   .
 \label{eq5}
\end{eqnarray}
It is also noted here that the deformation-induced migration in the form of eq. (\ref{eq1}) has been introduced in a different context of cell motility \cite{Casademunt}. In our model,  the coefficients $\kappa_2$ and $\kappa_3$ are positive assuming that the circular shape  is always stable when the deformation forces are absent. 
The  front-rear asymmetry is produced when the cell starts to migrate.

Deformations are caused by the internally created forces given by the terms  $F_{ij}^{(2)}$ and $F_{ijk}^{(3)}$  in eqs. (\ref{eq2}) and (\ref{eq3}),  respectively, which are called deformation forces. 
In the present model, these are force  moments defined on the cell boundary. Let us suppose a force $f_i(a, t)$ acting on the position $a$ on the boundary in a two-dimensional cell. The moments are defined by
\begin{eqnarray}
F_{ij}^{(2)}(t) &=&\int da x_if_j(a, t)   ,
 \label{Fij}  \\
 F_{ijk}^{(3)}(t)&=&\int da x_ix_jf_k(a, t)   ,
\label{Fijk3}
\end{eqnarray}
where the integral runs over the cell boundary. 
In experiments of crawling cells, the distribution of traction forces between the cell membrane and the substrate has been measured \cite{Tanimoto}. To make a link between the traction forces and the deformation force tensors, one needs further information inside the cell such as the viscoelastic properties. In the present simple model approach, we do not make such a systematic reduction of degrees of freedom but simply assume that these force tensors have the same symmetry as that of the deformation tensors given by eqs. (\ref{S11}) - (\ref{W-}). For example, we redefine  $F_{ij}^{(2)}-(\delta_{ij}/2)F_{kk}^{(2)}$ as $F_{ij}^{(2)}$ to make the  tensor traceless. In a manner consistent with the expressions (\ref{S11}), (\ref{S12}), (\ref{W+}) and (\ref{W-}), we write the force tensors as
\begin{eqnarray}
F_{ij}^{(2)}&=&g^{(2)}(t)\cos2\Theta_{ij}^{(2)}   ,
 \label{Fij2}\\
F_{ijk}^{(3)}&=&g^{(3)}(t)\cos3\Theta_{ijk}^{(3)}   ,
 \label{Fijk}
 \end{eqnarray}
 where the amplitudes $g^{(2)}(t)$ and $g^{(3)}(t)$ will be specified later. 
There are  basically two choices of the  phase $\Theta^{(n)}$. One is  to equate it with the deformation directions 
consistently with eqs.  (\ref{S11}), (\ref{S12}), (\ref{W+}) and (\ref{W-});
 \begin{eqnarray}
\Theta_{11}^{(2)}&=&\theta_2   \label{Theta21}  ,   \\
\Theta_{12}^{(2)}&=&\theta_2 -\frac{\pi}{2}   \label{Theta22}    ,  \\
\Theta_{111}^{(3)}&=&\theta_3   \label{Theta23}   ,   \\
\Theta_{222}^{(2)}&=&\theta_3 +\frac{\pi}{2}   .
 \label{Theta24}
 \end{eqnarray} 
The other is to equate it with the velocity direction, i.e., 
\begin{eqnarray}
\Theta_{11}^{(2)}&=&\theta   \label{Theta11}   ,  \\
\Theta_{12}^{(2)}&=&\theta -\frac{\pi}{2}  \label{Theta12}   ,\\
\Theta_{111}^{(3)}&=&\theta    \label{Theta13}   ,  \\
\Theta_{222}^{(2)}&=&\theta +\frac{\pi}{2}  ,
 \label{Theta14}
 \end{eqnarray} 
where the angle $\theta$ is related with the velocity of the center of mass as
\begin{eqnarray}
v_1&=& v\cos \theta  ,
 \label{v1}  \\
v_2&=&v\sin\theta  ,
 \label{v2}  
\end{eqnarray}
with the amplitude  $v>0$. 

 In the present paper, we mainly consider the case given by eqs. (\ref{Theta21})-(\ref{Theta24}).  In this case, the deformation forces are given by
\begin{eqnarray}
F_{11}^{(2)}&=&(g^{(2)}(t)+\xi_2(t) )\cos2(\theta_2+\eta_2(t))  ,
 \label{F11}\\
F_{12}^{(2)}&=&(g^{(2)}(t)+\xi_2(t) )\sin2(\theta_2+\eta_2(t))  ,
\label{F12}
\end{eqnarray}
\begin{eqnarray}
F_{111}^{(3)}&=&(g^{(3)}(t)+\xi_3(t) )\cos3(\theta_3+\eta_3(t))  ,
 \label{F111}\\
F_{112}^{(3)}&=&(g^{(3)}(t)+\xi_3(t) )\sin3(\theta_3+\eta_3(t))  ,
\label{F222}
\end{eqnarray}
where we have added random noises $\eta_2$, $\eta_3$, $\xi_2$ and $\xi_3$ both to the amplitudes and to the phases.
We assume that these noises obey the Gaussian statistics with zero means and are delta-correlated in time;
\begin{eqnarray}
<\eta_n>&=&<\xi_n> =0  ~~ (n= 2, 3)
\label{noise1i} \\
<\xi_n(t)\xi_m(t')>&=&\sigma_n^2 \delta_{nm}\delta(t-t')~~ (n= 2, 3)
\label{noise2} \\
<\eta_n(t)\eta_m(t')>&=&\zeta_n^2 \delta_{nm}\delta(t-t')~~ (n= 2, 3)
\label{noise3} \\
<\eta_n(t)\xi_m(t')>&=&0~ ~~~~~~~~~~~~~~~~~~(n= 2, 3)
\label{noise4} 
\end{eqnarray}
with $\sigma_n$ and $\zeta_n$ the magnitudes of the noises. 
We do not  add a random force to eq.  (\ref{eq1}) but explore the consequence that randomness of a migrating cell is generated by the stochasticity of the shape dynamics.

Equation (\ref{eq1}) can be written as
\begin{eqnarray}
 v_{1} &=&  \gamma[(S_{11}-S_{22})U_{111} -2S_{12}U_{222}]   \nonumber \\
 &=&\gamma[2s_2s_3\cos (3\theta_3-2\theta_2)]  ,
 \label{v11} \\
v_{2} &=&  \gamma[(S_{22}-S_{11})U_{222} -2S_{12}U_{111}] \nonumber \\
 &=&\gamma[2s_2s_3\sin (3\theta_3-2\theta_2)]  ,
 \label{v22} 
\end{eqnarray}
where $s_2=2a_2>0$ and $ s_3=2a_3>0$. Using eqs. (\ref{v1}) and (\ref{v2}), we obtain
\begin{eqnarray}
v&=&2|\gamma| s_2s_3   ,
 \label{vpai}\\
 \theta&=&3\theta_3-2\theta_2-\Psi_v   ,
 \label{thetapai}
\end{eqnarray}
where 
\begin{eqnarray}
\Psi_v&=&0  ,
 \label{Psi1}
 \end{eqnarray}
   for $\gamma>0$ and 
\begin{eqnarray} 
\Psi_v&=& \pi   ,
 \label{Psi2}
\end{eqnarray}
for  $\gamma<0$, 

Equations  (\ref{eq2}) and (\ref{eq3})  are written in terms of the amplitudes and angles of deformations as
\begin{eqnarray}
\frac{d s_{2}}{dt}&=&-\kappa_2 s_{2} +\frac{b_0v^2}{2}\cos(6\theta_{23}+2\Psi_v) \nonumber \\
&+&g^{(2)}(t)  +\xi_2(t)   ,
 \label{eqs22noise}  \\
 \frac{d \theta_{2}}{dt}&=&-\frac{b_0v^2}{4s_2}\sin(6\theta_{23}+2\Psi_v)
 +\frac{g^{(2)}(t)}{s_2}\eta_2(t)  ,
\label{eqtheta23noise} \\
\frac{d s_{3}}{dt}&=&-\kappa_3 s_{3}  + \frac{d_0v^3}{4}\cos(6\theta_{23}+3\Psi_v)  \nonumber \\
&+&g^{(3)}(t) +\xi_3(t)  ,
 \label{eqs32noise}  \\
\frac{d \theta_{3}}{dt}&=&-\frac{d_0v^3}{12s_3}\sin(6\theta_{23}+3\Psi_v)
+\frac{g^{(3)}(t) }{s_3}\eta_3(t)  ,
\label{eqtheta33noise}
\end{eqnarray}
where the relation (\ref{thetapai}) has been used. 
 Since the system is isotropic, the relative angle defined by
\begin{eqnarray}
\theta_{23}=\theta_2-\theta_3   ,
\label{theta23}
\end{eqnarray}
enters in the time-evolution equations.
We have ignored the terms nonlinear in the noises  employing the approximations such that $(\cos2\eta_2) \to 1$, $ (\cos3\eta_3) \to 1$ and $\xi_3(t)(\sin3\eta_3) \to 0$.

Now we discuss the amplitudes of the deformation forces $g^{(2)}(t) $ and $g^{(3)}(t) $.
When our model is 
applied to keratocyte cells which migrate coherently without irregular shape deformations, the deformation forces are assumed to be constant.  On the other hand, experiments of  ${\it Dictyostelium}$ cells show that the shape changes repeatedly such that a cell elongates first and then the third mode deformation follows \cite{Tanimoto}. Therefore,  the deformation forces in this case are time-dependent with a suitable phase difference between the modes. 
This means that if  the forces are approximated by periodic functions with period $T$ such that
\begin{eqnarray}
g^{(2)}(t)&=&g^{(2)}(t+T)   ,
\label{g2t} \\
g^{(3)}(t-\Phi)&=&g^{(3)}(t+T-\Phi)   ,
\label{g3t}
\end{eqnarray}
the phase difference $\Phi$ should be $0<\Phi < \pi$.

Another possibility is that generation of  the deformation forces has an excitable origin. As mentioned in Introduction, Sawai and his coworkers have found experimentally that motility of  a ${\it Dictyostelium}$ cell is coupled with chemical reactions inside the cell and they have analyzed the dynamics by a set of excitable reaction diffusion equations with noises \cite{Taniguchi}. In the present model approach, we utilize the so-called coherence resonance \cite{Coherence} to creat stochastic but almost periodic deformation forces. The phase difference between the modes as eq. (\ref{g3t}) is also considered as shown in section 4. 

From the solutions of eqs. (\ref{v11}), (\ref{v22}), (\ref{eqs22noise}), (\ref{eqtheta23noise}), (\ref{eqs32noise}), and (\ref{eqtheta33noise}), 
the motion of  a migrating cell is represented as
\begin{eqnarray}
x(t)&=&x_{cm}(t)+r(t)\cos (\alpha)    ,
\label{xcell} \\
y(t)&=&y_{cm}(t)+r(t)\sin (\alpha)    ,
\label{ycell} 
\end{eqnarray}
where  $0 \le \alpha < 2\pi$ and 
\begin{eqnarray}
r(t)=R_0\Big(1+s_2\cos2(\alpha-\theta_2)+s_3\cos3(\alpha-\theta_3)\Big)   ,
\label{rcell}
\end{eqnarray}
 with  $(x_{cm}, y_{cm})$ the position of the center of mass of the cell.

\section{Numerical results for constant deformation forces}

Hereafter throughout present paper, we consider the case $\gamma<0$ and $\Psi_v=\pi$ (eq. \ref{Psi2}) since this condition is found to correspond to motility of keratocyte cells and ${\it Dictyostelium}$ cells.

\begin{figure}
\begin{center}
\includegraphics[width=1.0\linewidth]{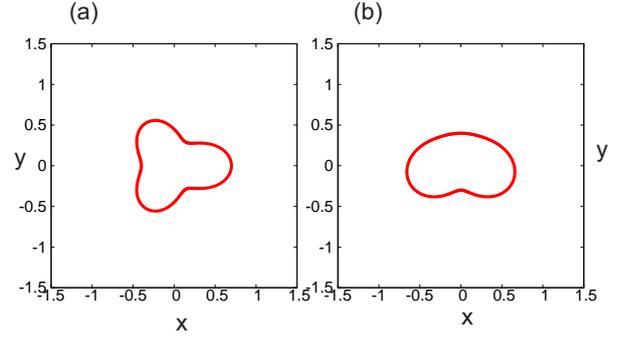}
\caption{Stationary shapes of a crawling cell subjected to constant deformation forces for (a) $R_0=0.5$, $s_2=0.1$, $s_3=0.3$ and  $\theta_2=\theta_3=0$, 
and (b)  $R_0=0.5$, $s_2=0.3$, $s_3=0.1$, $\theta_2=0$ and $\theta_3=-\pi/6$. When $\gamma<0$, the cell migrates to the left in (a)
and upward in (b). 
}
\label{constforce}
\end{center}
\end{figure}

In this section, we analyze the case that  the deformation forces are independent of time and noises are absent. 
From eqs. (\ref{eqs22noise}),  (\ref{eqtheta23noise}), (\ref{eqs32noise}) and (\ref{eqtheta33noise}) ignoring the noise terms, we obtain 
\begin{eqnarray}
\frac{d s_{2}}{dt}&=&-\kappa_2 s_{2} +\frac{b_0v^2}{2}\cos(6\theta_{23}) +g^{(2)}   ,
 \label{eqs22}  \\
\frac{d s_{3}}{dt}&=&-\kappa_3 s_{3} - \frac{d_0v^3}{4}\cos(6\theta_{23}) +g^{(3)}   ,
 \label{eqs32}  
\end{eqnarray}
\begin{eqnarray}
 \frac{d  \theta_{23}}{dt}&=&-C\sin(6 \theta_{23})   ,
\label{Delta} 
\end{eqnarray}
where
\begin{eqnarray}
C=\frac{b_0v^2}{4s_2}+ \frac{ d_0v^3}{12s_3}   .
\label{C} 
\end{eqnarray}
We require that the amplitudes $s_2$ and $s_3$ are positive. This condition is fulfilled for the stationary solutions of  eqs. (\ref{eqs22}) and (\ref{eqs32}) with eq.   (\ref{vpai}) if $|\gamma|$ is not very large. We put $\theta_2=0$ in eq. (\ref{rcell}) without loss of generality. That is, the long axis of an elliptical deformation is parallel to the $x$ axis. 
If $C$ is positive which is realized for $b_0>0$ and $d_0>0$, $\theta_{23}=0$ or $\pi/3$  are a stable solution. The former gives us $\theta_3=0$ and $\theta=-\pi$. The stationary shape in this case is displayed in Fig. \ref{constforce}(a). The cell is migrating to the left at a constant speed. The other solution $\theta_{23}=\pi/3$ gives us $\theta_3=-\pi/3$ and $\theta=-2\pi$. Therefore the motion and the shape in this case is just mirror symmetric with respect  to the $y$ axis of Fig.  \ref{constforce}(a).
When  $C$ is negative as is realized by the condition that $b_0<0$ and $d_0<0$, the stable solution of eq. (\ref{Delta}) is given by $\theta_{23}=\pm \pi/6$.
 The shape for $\theta_{3}=- \pi/6$ is shown in  Fig. \ref{constforce}(b).   The angle of the velocity direction is given by $\theta=-3\pi/2$ indicating that  the cell migrates upward. 
  The other case $\theta_{3}= \pi/6$ is just mirror-symmetric with respect to the $x$ axis. 
 The steady motion of keratocyte cells is qualitatively expressed as  in Fig.  \ref{constforce}(b) whereas a shape of  ${\it Dictyostelium}$ cells more resembles to that in Fig.  \ref{constforce}(a). When $\gamma$ is positive, the migration direction is reversed without any influence for the cell shape.

\begin{figure}[t]
\begin{center}
\includegraphics[width=0.8\linewidth]{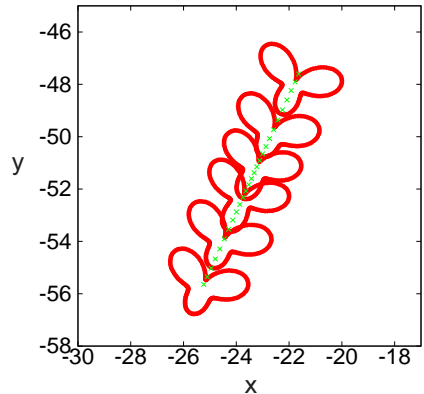}
\caption{Snapshots of a cell undergoing an oscillatory rectilinear motion from the top (at $t=175$) to the bottom (at $t=200$).  The parameters are set as   $\gamma=-1.0$, $\kappa_2=1$, $\kappa_3=0.12$, $b_0=d_0=-3.5$ and $g^{(2)}=g^{(3)}=0.1$. The symbols $\times$ indicate the trajectory of the center of mass.
}
\label{snapkerato}
\end{center}
\end{figure}

\begin{figure}[t]
\begin{center}
\includegraphics[width=0.7\linewidth]{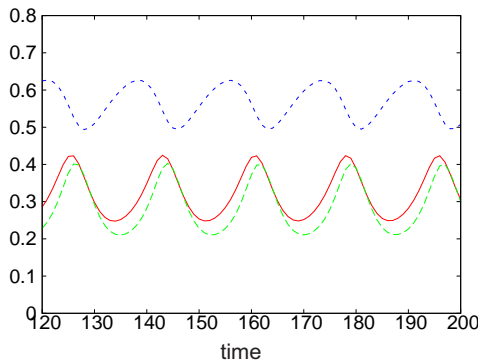}
\caption{Time-evolution of $v$ (red and solid line),  $s_2$ (green and  broken line) and $s_3$ (blue and dotted line). The parameters are the same as those in Fig. \ref{snapkerato}.
}
\label{evokerato}
\end{center}
\end{figure}

We investigate the dynamics of a cell in Fig. \ref{constforce}(b) for negative values of $b_0$ and $d_0$ in further detail. Let us introduce $\delta_{23}$ through the relation that 
\begin{eqnarray}
\theta_{23}=\frac{\pi}{6}+\delta_{23}.
\label{d23} 
\end{eqnarray}
Equations (\ref{eqs22}),  (\ref{eqs32}) and (\ref{C})  hold after replacement of $\theta_{23}$ by $\delta_{23}$, $b_0$ by $|b_0|$ and $d_0$ by $|d_0|$. Note that $\delta_{23}=0$ is a stable solution. 
The nonlinear terms in eqs. (\ref{eqs22}) and (\ref{eqs32}) have an interesting structure. 
Note that $v$ is proportional both to $s_2$ and to $s_3$. 
If $s_3$ is large, it makes $s_2$ increase in eq.  (\ref{eqs22}). This, in turn,  tends to decrease $s_3$ by the second term in eq.  (\ref{eqs32}). However, when the time-variation of $s_3$ is sufficiently slow, i.e., $\kappa_3 \ll \kappa_2$, there is a delay of this evolution and an overshooting  can occur, which make an oscillatory dynamics of $s_2$ and $s_3$ for constant values of $g^{(2)}$ and $g^{(3)}$. We have verified numerically that this is indeed the case. A supercritical bifurcation occurs for $\gamma=-1.0$, $\kappa_2=1$,  $b_0=d_0=-3.5$ and $g^{(2)}=g^{(3)}=0.1$
at around $\kappa^c_3 \approx 0.125$ below which the stationary solution of eqs. (\ref{eqs22}) and (\ref{eqs32}) looses stability and a limit cycle oscillation emerges. Figure \ref{snapkerato} displays snapshots of a migrating cell for $\kappa_3=0.12$  in the interval $175 \le t \le200$. The corresponding time-variations of $v$, $s_2$ and $s_3$ are shown in Fig. \ref{evokerato}. The period of oscillation $T$ is about $T=17$.

The oscillatory straight motion in Fig. \ref{snapkerato} should be compared with the one obtained in the phase field model of  keratocyte cells \cite{Ziebert1}. Figure 2 in ref. \cite{Ziebert1} indicates that the aspect ratio $h$ of a cell takes a maximum just after the velocity becomes a maximum. The asymmetry parameter $\zeta$ exhibits approximately an anti-phase oscillation with  the aspect ratio. See ref.  \cite{Ziebert1} for the definitions of $h$ and $\zeta$.  If we assume the correspondence  $h-1 \leftrightarrow s_2$ and  $\zeta  \leftrightarrow s_3$, which is valid for $s_2, s_3 \ll 1$, the dynamical  behavior in Fig. \ref{evokerato} is quite similar to  the one in ref.  \cite{Ziebert1}. Only the difference is that $s_3 > s_2$ in our model whereas 
$h-1 > \zeta$ in the phase field model so that the shape of an oscillating cell looks slightly different. It is also noted that the phase difference between $s_2$ and $s_3$ depends on the model parameters. For example, for   $\gamma=-1.0$, $\kappa_2=1.0$, $\kappa_3=0.05$,  $b_0=-4.0$, $d_0=-3.0$, $g^{(2)}=0.12$ and $g^{(3)}=0.06$, the amplitude $s_2$ reaches  the maximum before $s_3$ takes its minimum value. 

The oscillatory rectilinear motion occurs also for the parameters  of Fig.  \ref{constforce}(a) where the cell elongates parallel to the migration direction.  It is readily shown that when $\gamma<0$, and $b_0$ and $d_0$ are positive, the same argument of oscillation as the above can be applied to eqs. (\ref{eqs22}) and (\ref{eqs32}) with $\theta_{23}=\delta_{23}$. Therefore the phase difference between $s_2$ and $s_3$ is approximately equal to $\pi$. 
However, experiments of ${\it Dictyostelium}$ cells indicate that a migration cell tends to first elongate along the velocity direction and, shortly after this, grows a Y-shape deformation \cite{Tanimoto}. This fact is not in accord with the phase difference between $s_2$ and $s_3$ mentioned above. To achieve the experimentally  observed sequence of the shape change, we need to introduce time-dependent deformation forces as described in the next section. 
  
Finally we make a remark that if we add the fourth mode, $ c_{\pm4}(t)$ in the time-evolution equations, a circular motion appears as a stable solution after destabilization of a straight motion, but, other oscillatory motions such as a bipedal motion and a zig-zag motion have not been obtained at present.

\section{Numerical simulations for time-dependent deformation forces}

In this section, we put $\gamma <0$, $b_0>0$ and  $d_0>0$ to investigate the dynamics of a migrating cell as shown in Fig.  \ref{constforce}(a) under the influence of the time-dependent deformation forces and noises. 
First, we need to create the forces $g^{(2)}(t)$ and $g^{(3)}(t)$. Experiments of  ${\it Dictyostelium}$ cells indicate that the force dipole and the force quadrupole  are time-dependent but not precisely periodic  \cite{Tanimoto}.  The phase difference between these two is another  important parameter in constructing a model system. 
Furthermore, Sawai and his coworkers have interpreted the motility of ${\it Dictyostelium}$ cells as an excitable system \cite{Taniguchi}. In order to take this into account and to realize an appropriate phase difference in a stochastic excitable system, we employ the coherence resonance \cite{Coherence}.  That is, we consider the following excitable system;
\begin{eqnarray}
\tau\frac{d w}{dt}&=&\sigma(w-\frac{1}{3}w^3-z)   ,
\label{w} \\
\frac{d z}{dt}&=&\sigma(w+a+\epsilon_z)   ,
\label{z}
\end{eqnarray}
where $\sigma$ is a positive constant which is needed to adjust the average period of the deformation forces. 
The noise $\epsilon_z$ is Gaussian with 
\begin{eqnarray}
<\epsilon_z>&=&0   ,
\label{znoise1} \\
<\epsilon_z(t) \epsilon_z(t') > &=&d^2\delta(t-t') .
\label{znoise2}
\end{eqnarray}
where  $d$  is the noise intensity.

Here, we briefly summarize our method of numerical 
 computation of the stochastic equations eqs. (\ref{eqs22noise}),  (\ref{eqtheta23noise}), (\ref{eqs32noise}),  (\ref{eqtheta33noise}) and
   (\ref{z}), all of which take the following  form
\begin{eqnarray}
\frac{d X}{dt}&=&f(X)+h(X)z ,
\label{X}
\end{eqnarray}
where the Gaussian noise $z$ satisfies 
\begin{eqnarray}
<z>&=&0   ,
\label{Xnoise1} \\
<z(t)z(t') > &=&D_X^2\delta(t-t') .
\label{Xnoise2}
\end{eqnarray}
with the noise intensity $D_X$.
 We employ the standard Euler-Maruyama scheme \cite{Maruyama} for the discretized version as
  \begin{eqnarray}
X_{n+1}-X_n=\Delta tf(X_n)+(\Delta t)^{1/2} D_XG_nh(X_n),
\label{Xn}
\end{eqnarray} 
where $G_n$ is Gaussian random number with $<G_n>=0$ and    $<G_n^2>=1$, and $\Delta t$ the time increment. For simplicity, we replace $G_n$ by  another random number $\sqrt{3}G'_n$ where $G'_n$ is distributed uniformly between -1 and 1. It has been proved mathematically that numeral accuracy is unaltered by this replacement for sufficiently small $\Delta t$ \cite{Greiner}.

In eq. (\ref{z}), we put $d=1/(10\sqrt{3})$.
When $ a>1$, the solution $w=-a$, and $z=-a +a^3/3$ is linearly stable and the system is excitable when $\tau$ is very small.  
We have examined the dynamics in the regions  $1.01<a<1.1$ and $10<1/\tau<100$. For larger values of $a$ and for smaller values of $1/\tau$, the time interval between two adjacent excitations increases, but the duration time  of each excitation is insensitive to these parameters.

\begin{figure}
\begin{center}
\includegraphics[width=0.7\linewidth]{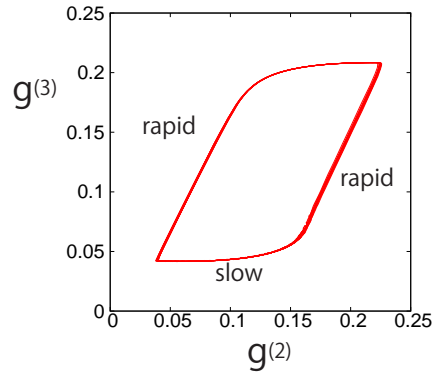}
\caption{Trajectory on the phase space $g^{(2)}$-$g^{(3)}$.
 }
\label{wzmod}
\end{center}
\end{figure}
\begin{figure}
\begin{center}
\includegraphics[width=0.7\linewidth]{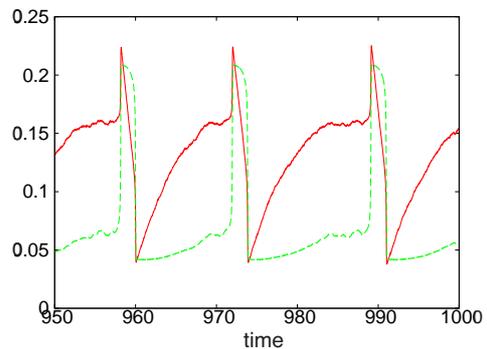}
\caption{Time-evolution of $g^{(2)}$ (red full line) and $g^{(3)}$ (green broken line).
}
\label{wzt}
\end{center}
\end{figure}

The deformation  forces are assumed to be given by
\begin{eqnarray}
g^{(2)}=g^{(2)}_1(w/3-2\sqrt{2}z/3+g^{(2)}_2)    ,
\label{g22} \\
g^{(3)}=g^{(3)}_1(2\sqrt{2}w/3+z/3+g^{(3)}_2)   ,
\label{g32}
\end{eqnarray}
where 
 we put  $g^{(2)}_1=0.07$, $g^{(2)}_2=2$, $g^{(3)}_1=0.05$ and $g^{(3)}_2=2.5$ unless stated otherwise.
The behavior of  $g^{(2)}$ and $g^{(3)}$ for $\tau=0.01$, $a=1.05$ and $\sigma=0.3$ is shown in Figs. \ref{wzmod} and \ref{wzt}. Note that $g^{(2)}$ increases  first while $g^{(3)}$ remains small and  almost constant, and then both exhibit a rapid increase. After reaching  the maximum, the forces decrease  rapidly and the process is repeated. Note also that this dynamics  is noise-controlled and is not strictly periodic as can be seen in Fig. \ref{wzt}.

\begin{figure}
\begin{center}
\includegraphics[width=0.8\linewidth]{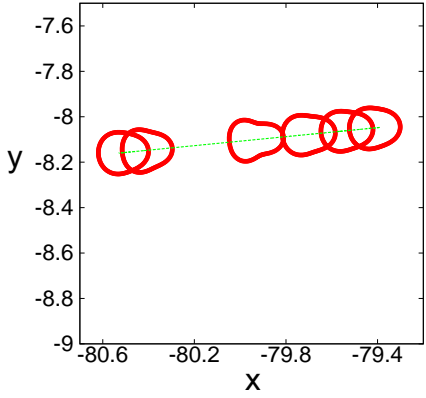}
\caption{Snapshots of a cell migrating to the left in the absence of noises  at $t=966.25$,   $t=968.75$, $t=971.25$, $t=973.75$, $t=975.25$ and $t=977.75$. The straight line is the trajectory of the center of mass.
}
\label{shapenonoise}
\end{center}
\end{figure}

We have solved numerically eqs.  (\ref{v11}), (\ref{v22}), (\ref{eqs22noise}), (\ref{eqtheta23noise}), (\ref{eqs32noise}), and (\ref{eqtheta33noise}) with the forces (\ref{g22}) and (\ref{g32}). 
As mentioned above, the Euler-Maruyama  scheme with the  time increment  $\Delta t=10^{-4}$ is used in the presence of noises.
Figure~\ref{shapenonoise} displays a migrating cell without noises $\sigma_{n}=\eta_{n}=0$ for the parameters  $\gamma=-4$, $R_0=0.1$, $b_0=d_0=1.0$, $\kappa_2=\kappa_3=1.0$. 
An almost circular  cell first elongates in the direction of the migration velocity, then the third mode of deformation grows so that the cell is accelerated. As the deformations  become small, the cell gets slowdown and relaxes to the circular shape.

\begin{figure}
\begin{center}
\includegraphics[width=1.0\linewidth]{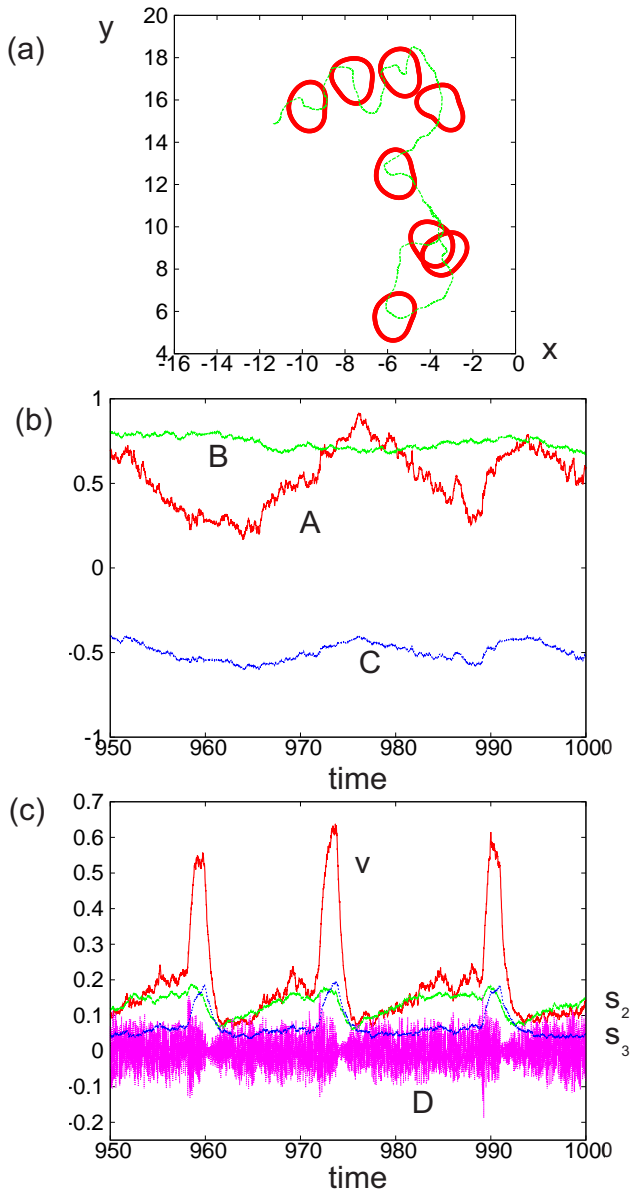}
\caption{(a) Trajectory  of the center of mass (green line) and the shape change (red lines) in a cell migration  for the linear model.  
The snapshots are taken  for $812.5 \le t \le 987.5$ with the time-interval $\delta t=25$. 
The cell starts to migrate around at $x=-4.0$ and $y=11$ at $t=800$. First it moves downward and then upward to $x=-11.4$ and $y=14.9$ at $t=1000$. 
(b) Time-evolution of $A=\theta/(2\pi)$ (red line), $B=\theta_2/(2\pi)$  (green line), and  $C=\theta_3/(2\pi)$ (blue line) for the linear model.
It is evident that the velocity direction $\theta$ exhibits a correlated time-change with  the angle of the third mode of deformations $\theta_3$. Such a correlation is less clear in the angle $\theta_2$ of the second mode. 
({c}) Time-evolution of $v$ (red line), $s_2$  (green line), $s_3$ (blue line)  and $D=\dot{\theta}\times 10^{-3}$ (purple line).
}
\label{linear}
\end{center}
\end{figure}

\begin{figure}
\begin{center}
\includegraphics[width=1.0\linewidth]{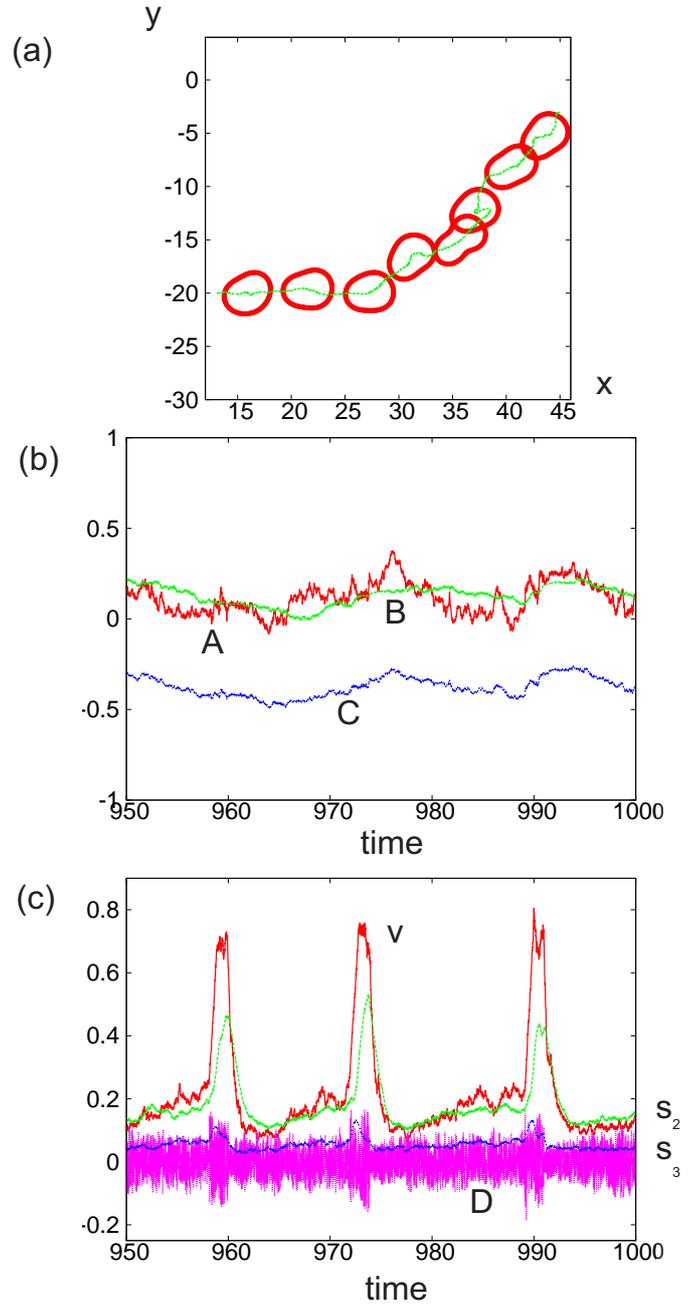}
\caption{(a) Trajectory  of the center of mass (green line) and the shape change (red lines) of a migrating (from the left to the right)  cell  with interval $\delta t=25$  in the presence of the nonlinearity. Note that the linear size of this figure is about twice as large as that of Fig. \ref{linear}(a).
(b)Time-evolution of $A=\theta/(2\pi)$ (red line), $B=\theta_2/(2\pi)$  (green line), and  $C=\theta_3/(2\pi)$ (blue line) for the nonlinear model.
It is evident that the velocity direction $\theta$ exhibits a correlated time-change with  the angles of the deformations $\theta_2$ and $\theta_3$. 
({c}) Time-evolution of $v$ (red line), $s_2$  (green line),  $s_3$ (blue line)  and $D=\dot{\theta}\times 10^{-3}$ (purple line).}
\label{nonlinear}
\end{center}
\end{figure}

Numerical simulations in the presence of random noises have been carried out both for  the linear case  $b_0=d_0=0$ and for the nonlinear case $b_0=d_0=2$.
Other parameters are fixed as $\gamma=-10$, $\kappa_2=\kappa_3=1.0$, 
$\epsilon_{amp}\equiv \sqrt{3}\sigma_n=0.02$ and $\epsilon_{ang}\equiv \sqrt{3}\eta_n=0.2$ ($n=2, 3$). 
Figure ~\ref{linear} shows the results for the linear case. The snapshots of the migrating cell with $R_0=1.0$  are shown in Fig.~\ref{linear}(a).  The time-evolutions of the angles $\theta$, $\theta_2$ and $\theta_3$ are depicted in Fig.~\ref{linear}(b) 
whereas those of $s_2$, $s_3$,  $v$  and $\dot{\theta}$ are 
in Fig.~\ref{linear}({c}).
The discretizing version of $\dot{\theta}=3\dot{\theta}_3-2\dot{\theta}_2$ as eq. (\ref{Xn}) divided by $\Delta t$  is used  in the plot of $\dot{\theta}$. The corresponding figures for the nonlinear case are given in Figs.  \ref{nonlinear}(a), (b) and ({c}), respectively.   The radius of the cell in  Fig. \ref{nonlinear}(a) is chosen as $R_0=2.0$ because the liner size of this figure is about twice as large as that of Fig. \ref{linear}(a).

Now we compare the two cases in detail. 
There are distinct differences in the dynamics between the linear and nonlinear cases. The trajectory of the linear case in Fig.  \ref{linear}(a)   is much more random compared with that in the nonlinear case in Fig. \ref{nonlinear}(a). In other words, a cell having the nonlinearity exhibits orientational persistence  in its migration. In the present set of parameters, the nonlinear terms make the migration velocity parallel to the elongation direction of the cell whereas the deformation directions and the migration direction are determined solely by the random noises as in eqs. (\ref{eqtheta23noise}) and (\ref{eqtheta33noise}) when the nonlinear terms are absent.  One unexpected behavior is that a cell reverses sometimes the direction of migration in the nonlinear case as can be seen around at $x=38$ and $y=-13$ in Fig.  \ref{nonlinear}(a)    although this does not occur for weaker noise intensities.  

One may note that the time change of the migration direction $\theta$ in Fig. \ref{linear}(b)  looks similar to that in Fig.  \ref{nonlinear}(b) apart from their amplitude. This is because these data were obtained by the same sequences of random noises. We have verified the random motion in the linear case and the persistent motion in the nonlinear case by other several independent runs of numerical simulations.

Although the motion in Fig.  \ref{linear}(a)  looks like a Brownian motion, the center-of-mass motion and the shape change are not totally independent of each other because of the relations (\ref{vpai}) and (\ref{thetapai}). It is evident in Fig.~\ref{linear}(b) that, in the linear case,  $\theta$ exhibits a time-variation  similar to $\theta_3$ but there is no apparent correlation between $\theta$ and $\theta_2$. In contrast, these three angles change in an almost synchronized manner in the nonlinear case as can be seen in Fig.  \ref{nonlinear}(b). Note that the variation of $\theta$ is in the range of $0<\theta<2\pi$  (modulus $2\pi$) so that the trajectory is curved substantially in the linear case whereas it is smaller in the nonlinear case as  $0<\theta<\pi$ (modulus $2\pi$) so that the trajectory does not exhibits much wandering.

The migration velocity is determined by the amplitudes $s_2$ and $s_3$ as eq. (\ref{vpai}). 
Note that the magnitude of the migration velocity is not much different between the linear and nonlinear cases. Therefore, the longer persistence length in the nonlinear case is not due to velocity difference. 
The two deformations $s_2$ and $s_3$ take a maximum almost simultaneously in the linear case as in Fig.  \ref{linear}({c})   because of the time-dependence of the forces as shown in Fig. \ref{wzt}. In  the nonlinear case depicted in Fig.  \ref{nonlinear}({c}), the amplitude $s_2$ becomes maximum  with a slightly delay compared with $s_3$.  The value of $s_3$ is much smaller than that of $s_2$.   This is a consequence of the fact that the nonlinear term in eq. (\ref{eqs22noise}) has a plus sign but that in eq. (\ref{eqs32noise}) has a minus sign for $\gamma<0$. 
The magnitude of fluctuations of $\dot{\theta}$ indicates an interesting time-variation as shown in Figs.  \ref{linear}({c}) and \ref{nonlinear}({c}). In the linear case, it becomes large just at the instance that the velocity (and $s_2$ and $s_3$) starts increasing. In the time region when these quantities are decreasing, the fluctuations of $\dot{\theta}$ are also small. The main reason of this is due to the multiplicative noises for the angles in eqs.  (\ref{eqtheta23noise}) and (\ref{eqtheta33noise}). Because of a delay in the evolutions of the deformations $s_2$ and $s_3$ relative to  the deformation forces, the cell is not deformed substantially even when the forces are large and vice verse in the linear case. In the nonlinear case, $\dot{\theta}$ fluctuates strongly during the growth of the velocity whereas appreciable decrease in fluctuations is not seen in the decreasing of the velocity since the nonlinearity masks the multiplicative effect of noises.  

We have assumed in this section that the origin of the deformation forces is excitability. The case of oscillating deformation forces will be described shortly in the next section. 

Before closing this section, we compare our results with experiments  of ${\it Dictyostelium}$ cells in a vegetative condition \cite{Tanimoto} to check the consistency of the characteristic space-time scale. 
The characteristic period of the force moments is about $2-3$ ~minutes. The migration (end-to-end) distance is about $40~\mu$m during 15 minutes. Since the diameter of a cell is about $15 ~\mu$m, this means that a cell migrates roughly  the distance of one cell size  in two cycles of deformations.  On the other hand, the average period of the deformation force in our model has been set about 15 in the unit of $\kappa_2=1$.  Figure ~\ref{linear}(a) displays the trajectory for the duration time 175 (about 11 cycles) during which the cell migrates at the distance about 10 (end-to-end distance). Since the radius is set as $R_0=1$, that is, the diameter is equal to 2, this is qualitatively consistent with the experiments.

\section{Discussion}

We have proposed a model system for cell crawling which is a set of nonlinear equations for deformations coupled through the migration velocity. We have introduced random noises for the time-evolution equations of deformations but not for the equation of the migration velocity since our basic assumption is that the internal forces cause crawling.  When the deformation forces are constant, we have found that the model has a limit cycle solution. This oscillatory behavior is compared qualitatively with that in the phase field model coupled with the polarization field for keratocyte cells \cite{Ziebert1}. 

In the case of time-dependent forces, they are generated by the coherence resonance. As a result, the deformations $s_2$ and $s_3$ exhibit a steep change in time. Because of the noise term $\epsilon_z$ in eq. (\ref{z}), the velocity $v$ and the phases $\theta_2$ and $\theta_3$ as well as $s_2$ and $s_3$ are stochastic. However, this stochasticity is not strong enough to make a random motion like in Fig.  \ref{linear}(a). Therefore we have introduced another set of random  noises $\xi_n$ and $\eta_n$ as eqs. (\ref{F11}) - (\ref{F222}) in the time-evolution equations. This is a technical point in the present theory. We do  not claim that  two different kinds of noises are involved in the motility of real living cells. 

For milder time-dependence, one may introduce oscillatory (non-stochastic) deformation forces such as
\begin{eqnarray}
g^{(2)}(t)=g^{(2)}_c+g^{(2)}_0\left(\frac{1+\cos (\omega t)}{2}\right)^2     ,
\label{g2tmodify} \\
g^{(3)}(t)=g^{(3)}_c+g^{(3)}_0\left(\frac{1+\cos (\omega t-\Phi) }{2}\right)^2    ,
\label{g3tmodify}
\end{eqnarray}
where $g^{(2)}_c$, $g^{(2)}_0$, $g^{(3)}_c$ and  $g^{(3)}_0$  are positive constant.  
We have carried out numerical simulations of eqs.  (\ref{v11}), (\ref{v22}), (\ref{eqs22noise}), (\ref{eqtheta23noise}), (\ref{eqs32noise}), and (\ref{eqtheta33noise}) with (\ref{g2tmodify}) and (\ref{g3tmodify})  with either $\Phi=\pi/4$ or $\Phi=\pi/2$, and $0 \le \omega \le 2\pi$.
The results both for the linear case and the nonlinear case are not much different qualitatively from those in the excitable deformation forces shown in the preceding section. By utilizing the oscillatory deformation forces in the present model, the motility of human hematopoietic stem cells is now under investigation \cite{Tanaka2}.

We have emphasized that the phase difference of the deformation forces and the relative angle of the dynamical variables are crucial to compare our model with experiments. For example, we have chosen $\Psi_v=\pi$ throughout the present paper. Anti-phase oscillation of $s_2$ and $s_3$ for the constant deformation forces seems to be consistent with the motility of keratocyte obtained theoretically \cite{Ziebert1} as shown in section 3.  In section 4, the time-dependence of the deformation forces are chosen such that the  value of  $(g^{(2)}, g^{(3)})$ moves counter-clockwise as shown in Fig. \ref{wzmod}, or the phase $\Phi$ in eq. (\ref{g3tmodify}) is $0<\Phi\le \pi$.  

We have chosen the direction of deformation forces as eqs. (\ref{Theta21})-(\ref{Theta24}). It is mentioned here briefly that what happens if one uses the relations (\ref{Theta11})-(\ref{Theta14}). In this case, equations for the amplitudes and the phases of deformations are given by
\begin{eqnarray}
\frac{d s_2}{dt}&=&-\kappa_2 s_2 +g^{(2)}(t)\cos (6\theta_{23}+2\Psi_v)   ,
\label{eqamp2} \\
\frac{d \theta_2}{dt}&=&-\frac{g^{(2)}(t)}{2s_2}\sin (6\theta_{23}+2\Psi_v)   ,
\label{eqphase2} \\
\frac{d s_3}{dt}&=&-\kappa_3 s_3 +g^{(3)}(t)\cos (6\theta_{23}+3\Psi_v)   ,
\label{eqamp3} \\
\frac{d \theta_3}{dt}&=&-\frac{g^{(3)}(t)}{3s_3}\sin  (6\theta_{23}+3\Psi_v)   .
\label{eqphase3} 
\end{eqnarray}
For the sake of clarity, we have written down the equations for the linear case without noises. If one sets $\Psi_v=\pi$, the stationary solution of the amplitude $s_3$ turns out to be negative since $\kappa_3$ and $g^{(3)}(t)$ are assumed to be positive. If $\Psi_v=0$, i.e., $\gamma>0$, this difficulty is removed, but in this case, the cell takes the same shape as  in Fig. \ref{shapenonoise}, but  migrates from the left to the right with the $s_2$ deformation followed by the $s_3$ deformation. Such a motion, however, has not been observed  experimentally in ${\it Dictyostelium}$ 
cells. See, e.g., ref.  \cite{Tanimoto}.

We have shown that the correlation between the deformations and the migration velocity plays an important role in the cell crawling. First of all, it is noted that 
the relation (\ref{thetapai}) causes  a  correlation between them
even when the nonlinear coupling is absent. Secondly, when the nonlinearity is present, the motion becomes more persistent. That is, the cell tends to migrate  without large changes of the direction although  the intensity of the random noises is the same as in the linear case. We note that Maeda et al have observed  a similar  dynamics  experimentally in   ${\it Dictyostelium}$ cells \cite{Maeda}. They have made a systematic comparison of the motility in a starved condition and  a vegetative condition. A cell in a  starved condition migrates at a larger velocity and elongates in the direction parallel to the migration direction. Its trajectory is much more persistent than in a  vegetative condition. In the present analysis, we have changed the coefficients of the nonlinear terms to compare the linear and nonlinear cases explicitly. It is worth mentioning, however, that the nonlinear effects are enhanced for larger migration velocity by increasing $|\gamma|$ in eq. (\ref{vpai}) keeping other parameters fixed.  It is an interesting future problem to  elucidate further both theoretically and experimentally the difference of the dynamical behavior between the starved and vegetative conditions. 

In the present study, we have considered only the second and third mode deformations assuming that higher modes relax sufficiently rapidly and the forces acting on them  are weaker. 
Preliminary numerical investigations have been  carried out  by including the fourth mode with the relaxation constant comparable with those of the second and third modes. 
In this case, as mentioned at the end of section 3,  circular motions appear for time-independent forces in the absence of noises. On the other hand, when the forces of the fourth mode are time-dependent  as eqs. (\ref{g2tmodify}) and (\ref{g3tmodify}), spinning motion,  quasi-periodic motion and chaotic motion have been found. 
We do not go into detail here but will publish these results separately elsewhere. 

Obviously, a remaining problem is to derive the time-evolution equations (\ref{eq1}), (\ref{eq2}), and  (\ref{eq3}) starting either from the phase field model of cell crawling, which involves the traction force explicitly \cite{Ziebert2} or from the model of active droplets subjected to active stress \cite{Tihung}  by means of the reduction theory as was done for traveling domains in reaction-diffusion systems \cite{OOS} and for hydrodynamic droplets self-propelled by the Marangoni effect \cite{Yabunaka}. However, this is left for a future study.

\subsection*{Acknowledgments}
We would like to thank Satoshi Sawai for valuable discussions.
This work was supported by Grant-in-Aids for Scientific Research A (No. 24244063)  from MEXT.

\end{document}